\documentstyle[preprint,aps]{revtex}

\input psfig.sty
\tightenlines 

\begin{document}
\draft
\title{Plasma effects in neutrino-pair emission due to Cooper pairing of
protons in superconducting neutron stars\thanks{%
Contribution to the International Workshop ``HADRON PHYSICS 2002'', April 14
-- 19, Bento Gon\c{c}alves, Rio Grande do Sul, Brazil.}}
\author{L. B. Leinson}
\address{Institute of Terrestrial Magnetism, Ionosphere and Radio Wave\\
Propagation RAS, 142190 Troitsk, Moscow Region, Russia\\
E-mail: leinson@izmiran.rssi.ru}
\maketitle

\begin{abstract}
The neutrino emission due to formation and breaking of Cooper pairs of
protons in superconducting cores of neutron stars is considered with taking
into account the electromagnetic coupling of protons to ambient electrons.
Our calculation shows that the contribution of the vector weak current to
the $\nu\bar\nu$ emissivity of protons is much larger than that calculated by
different authors without taking into account the plasma effects. Partial
contribution of the pairing protons to the total neutrino radiation from the
neutron star core is very sensitive to the critical temperatures for the
proton and neutron pairing and can dominate in some domains of these
parameters.
\end{abstract}


\widetext

When the temperature inside a neutron star core is lower than the critical
temperature $T_{c}$ for nucleon pairing, the nucleon matter consists of a
condensate of Cooper pairs, which has thermal excitations in the form of not
paired quasi-particles. Cooper-pair formation and pair-breaking coexist in
statistical equilibrium and result in additional neutrino-pair emission from
the neutron star. Under certain conditions, neutrino emission due to Cooper
pairing of nucleons can dominate in neutrino energy losses from the neutron
star. The mechanism of neutrino emission due to the singlet-state pairing of
neutrons was proposed by Flowers et al. \cite{FRS76} many years ago.

Specifics of neutrino emission due to the proton pairing occurs because of a
smallness of the vector weak coupling of a proton. Cooper pairing of protons
takes place likely in $^{1}S_{0}$-state \cite{TT93}. When protons are
treated non-relativistically, the total spin of the Cooper pair in the
singlet-state is zero. By this reason, the axial-vector contribution of the
proton weak current to the neutrino emissivity occurs only as a relativistic
correction. Therefore, neutrino emission produced by the proton pairing is
conventionally estimated negligible \cite{KHY99}.

Such inference is made on the basis of calculations which ignored
electromagnetic correlations among the charged particles in the QED plasma.
Actually, protons are coupled to ambient electrons via the electromagnetic
field. By undergoing a quantum transition to the paired state, protons
polarize the medium, thus inducing the motion of electrons inside the Debye
sphere around them. The electron weak current associated to this motion
generates neutrinos coherently with the weak current of protons, because the
wavelength $\lambda $ of radiated neutrino pairs is much larger than the
electron Debye screening distance $D_{e}$ (typically, $D_{e}^{2}/\lambda
^{2}\sim 10^{-2}$). The induced radiation from ambient electrons many times
exceeds the neutrino radiation from the initial proton pair \cite{L00}, \cite%
{L01}. To demonstrate this we study the neutrino emissivity, caused by the
proton pairing, by the use of the fluctuation-dissipation theorem. We
consider the total energy which is emitted into neutrino pairs per unit
volume and time. For one neutrino flavor, the emissivity is given by the
following formula: 
\[
Q=\frac{G_{F}^{2}}{2}\int \;\frac{\omega }{\exp \left( \frac{\omega }{T}%
\right) -1}2\mathop{\rm Im}\tilde{\Pi}_{R}^{\mu \nu }\left( K\right) %
\mathop{\rm Tr}\left( j_{\mu }j_{\nu }^{\ast }\right) \frac{d^{3}k_{1}}{%
2\omega _{1}(2\pi )^{3}}\frac{d^{3}k_{2}}{2\omega _{2}(2\pi )^{3}}, 
\]%
where the integration goes over the phase volume of neutrino and
antineutrino of the total energy $\omega =\omega _{1}+\omega _{2}$ and the
total momentum ${\bf k=k}_{1}+{\bf k}_{2}$. In this formula $G_{F}$ is the
Fermi coupling constant, $\tilde{\Pi}^{\mu \nu }$ is the retarded
polarization tensor of the medium which has ends at the weak vertex, and the
neutrino weak current is of the standard form 
\begin{equation}
j_{\mu }=\bar{\nu}\gamma _{\mu }\left( 1-\gamma _{5}\right) \nu .  \label{14}
\end{equation}

To incorporate the collective plasma effects, one should take into account
exchange of photons between charged particles in the plasma. This can be
done in the Random phase approximation (RPA). In this case the weak
polarization tensor of the medium is given by the diagrams shown in Fig. 1.

\vskip0.3cm \psfig{file=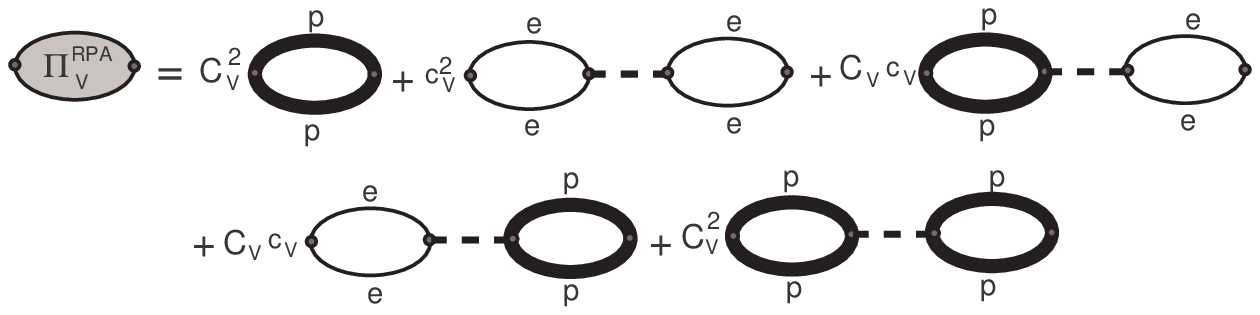} Fig. 1. RPA approximation to the weak 
polarization of the medium. Here $C_{V}=0.04$ is the constant of the proton 
vector weak coupling with the
neutrino field; $c_{V}$ is the electron coupling constant, ($c_{V}=0.96$ for
electron neutrinos, and $c_{V}^{\prime}=-0.04$ for muon and tau neutrinos).
\vskip0.3cm\noindent
Here the first term describes the contribution, which comes directly from the 
pairing protons, as shown in Fig. 2.
 
\vskip0.3cm \psfig{file=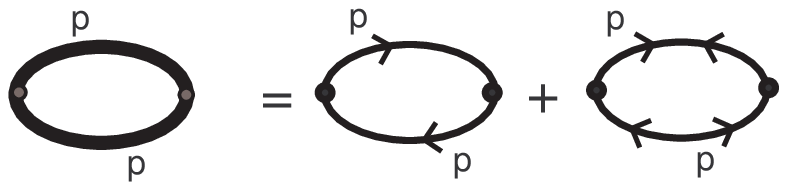}Fig. 2. Free-gas approximation to the weak 
polarization of protons.
\vskip0.3cm\noindent
The other terms are caused by the plasma polarization.

To demonstrate efficiency of the collective effects, the $Q^{{\rm RPA}}$ is
plotted in Fig. 3 versus the dimensionless temperature $\tau =T/T_{c}$
together with that obtained in the free-gas approximation (FG) without
collective effects. One can see that, the collective effects substantially
enhance, the $\nu \bar{\nu}$ emissivity caused by the vector weak current of
protons.
\vskip0.3cm \psfig{file=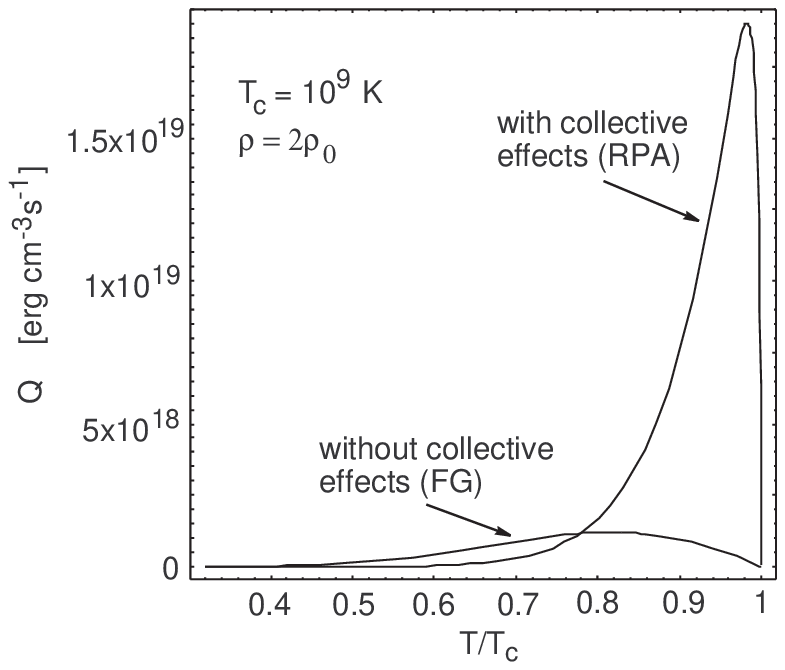}Fig. 3. Temperature dependence of the RPA vector weak current contribution
to the neutrino emissivity in comparision with that obtained in the free-gas
approximation (FG). The emissivities are shown versus the dimensionless
temperature $\protect\tau =T/T_{c}$ for beta-equilibrium nuclear matter.
\vskip0.3cm

The effect of neutron superfluidity and/or proton superconductivity on
different neutrino reactions in the neutron star core is very complicated.
Different neutrino production mechanisms can dominate at different cooling
stages depending on the temperature, and the matter density, and vary along
with the chosen parameters $T_{cp}$, $T_{cn}$, which are the critical
temperatures for the proton and neutron pairing. The Fig. 4 shows efficiency
of different reactions at the baryon density $\rho =2\rho _{0}$, where the
direct Urca process is forbidden. Neutrino emissivities due to the
singlet-state proton pairing, and the triplet-state neutron pairing are
plotted in logarithmic scale against the temperature together with the total
emissivity of two branches of the modified Urca processes, and the total
bremsstrahlung emissivity caused by nn-, np-, and pp-scattering, which are
suppressed due to the neutron superfluidity and/or proton superconductivity.
Two panels of Fig. 4 differ only by different choice of parameters $T_{cp}$
and $T_{cn}$. On the left panel we took $T_{cp}=5.6\times 10^{8}$ K, and $%
T_{cn}=5.6\times 10^{9}$ K. On the right panel we assume $T_{cp}=3.5\times
10^{9}$ K, and $T_{cn}=8.5\times 10^{8}$ K. Partial contributions of the
proton and neutron pairing to the total energy losses are very sensitive to
the corresponding critical temperatures.

\vskip0.3cm \psfig{file=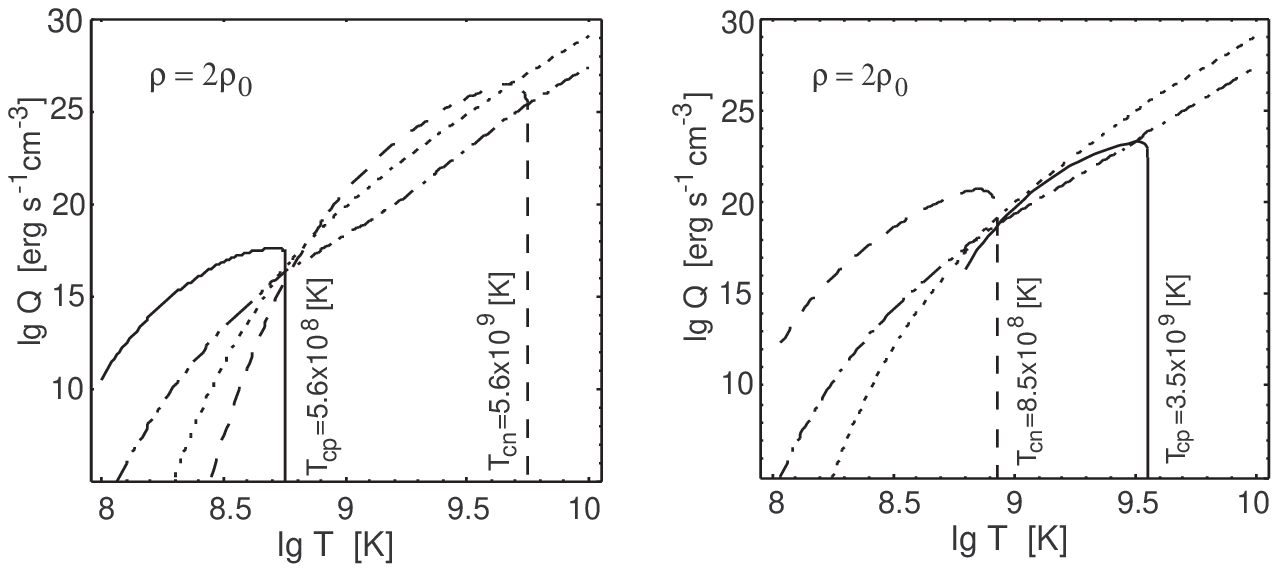}Fig. 4. Temperature dependence of the neutrino emissivity in different
reactions for $\protect\beta $-equilibrium nuclear matter of the density $%
\protect\rho =2\protect\rho _{0}$. The neutrino emissivity due to the
triplet-state neutron pairing is shown by dashed line. The solid line is the
neutrino emissivity due to the singlet-state proton pairing. Dot-and-dash
line shows the total bremsstrahlung emissivity caused by nn-, np-, and
pp-scattering; the dotted line exhibits the total emissivity of two branches
of the modified Urca processes. On the left panel we took $T_{cp}=5.6\times
10^{8}$ K, and $T_{cn}=5.6\times 10^{9}$ K. On the right panel we assume $%
T_{cp}=3.5\times 10^{9}$ K, and $T_{cn}=8.5\times 10^{8}$ K.

\end{document}